# Spatiotemporal Airy rings wavepackets

Xiaolin Su[1], Andy Chong[2,3], Qian Cao[1,4*] and Qiwen Zhan[1,4,5]

[1] School of Optical-Electrical and Computer Engineering, University of Shanghai for Science and Technology, Shanghai 200093, China.

[2] Department of Physics, Pusan National University, Busan 46241, Republic of Korea

[3] Institute for Future Earth, Pusan National University, Busan 46241, Republic of Korea

[4] Zhangjiang Laboratory, Shanghai 201210, China

[5] Westlake Institute for Optoelectronics, Fuyang, Hangzhou, China

*cao.qian@usst.edu.cn

**Abstract:**

Airy waves, known for their non-diffracting and self-accelerating properties, have been extensively studied in spatial and temporal domains, but their spatiotemporal (ST) counterparts remain largely unexplored. We report the first experimental realization of a spatiotemporal Airy rings wavepacket, which exhibits an Airy function distribution in the radial dimension of the ST domain. The wavepacket demonstrates abrupt autofocusing under the combined effects of diffraction and dispersion, achieving a 110 μm spatial and 320 fs temporal focus with a sharp intensity contrast along the propagation direction—ideal for nonlinear microscopy and multiphoton 3D printing. Notably, the wavepacket retains its autofocusing capability even after spatial obstruction, showcasing robust self-healing. Furthermore, by embedding a vortex phase, we create an ST-Airy vortex wavepacket that confines transverse orbital angular momentum (t-OAM) within a compact ST volume, enabling new avenues for studying light-matter interactions with t-OAM. Our findings advance the fundamental understanding of ST Airy waves and highlight their potential for transformative applications in ultrafast optics, structured light, and precision laser processing.

**Introduction**

Airy function, a mathematical solution that later became foundational in wave theory, was first introduced and named by the English mathematician Sir George Biddell Airy in 1838. In 1979, Berry and colleagues studied the Airy wave within the framework of the Schrödinger equation, revealing its remarkable properties, including non-diffraction and self-bending [1]. A perfect Airy beam, characterized by infinite energy, is a theoretical construct and cannot exist in practical systems [2, 3]. In 2007, Siviloglou et al. achieved a breakthrough by generating and observing finite-energy one-dimensional (1D) and two-dimensional (2D) optical Airy beams in the laboratory [4, 5]. This was made possible through the use of programmable devices like spatial light modulator (SLM), which enable precise control over the cubic phase modulation of Gaussian beams to create and manipulate Airy beams [6, 7]. Since these pioneering experiments, Airy beams have evolved into a versatile and practical tool in optics, driving extensive research into their properties and enabling a wide range of applications, including advanced imaging [8, 9], laser machining [10], optical tweezers [11, 12], and nonlinear optics researches [13]. Beyond the conventional Airy beam, other forms of optical Airy waves, such as Airy pulses [14, 15], symmetric Airy rings beam [16, 17],and electron Airy beams [18], have also been thoroughly investigated, further expanding the scope of their utility and understanding.

In recent years, the field of spatiotemporal optics, which explores light fields with a coupled spatial and temporal distribution, has garnered significant attention. This emerging area offers new opportunities for controlling light in ways that were previously unattainable, enabling unique photonic properties and phenomena [19-22] Despite these advancements, the study on the spatiotemporal form of optical Airy waves remains largely unexplored [23], representing a critical gap in the field. In this work, we present the first experimental realization of a spatiotemporal (ST) Airy rings wavepacket, where the Airy wave is structured radially in the spatiotemporal domain. The resulting ST-Airy rings wavepacket exhibits remarkable spatiotemporal abrupt auto-focusing and self-healing properties, making it highly promising and robust in applications where an abrupt increasing intensity contrast is needed, for example, nonlinear microscopy and multiphoton 3D printing. In the experiment, we also create the ST-Airy rings vortex wavepacket that can carry transverse orbital angular momentum (t-OAM) and is capable of confining t-OAM within a tightly focused spatiotemporal volume upon dispersive propagation. This might find application in studying light and matter interactions[24-26]. We believe these findings can advance the fundamental understanding of spatiotemporal Airy waves and highlight their potential for transformative applications in optics and photonics.

**Results**

Optical Airy wave has a finite energy and is generated by modulating the spectral light field with a third-order phase [27-29]. Figure 1A illustrate the profile of the perfect Airy wave with an infinite energy (upper one) and the optical Airy wave with a finite energy (lower one). The major difference is that optical Airy wave has a finite spectrum. To generate Airy beams as shown in Fig. 1B and C, the third-order phase is applied in the spatial frequency domain orthogonally/radially [4,5]. To generate the spatiotemporal (ST) Airy rings wavepacket, that is shown in Fig. 1D, we apply the radial third-order phase in the spatiotemporal Fourier domain, or in the $k_x - \omega$ plane of the input light field. Figure 1E shows the phase profile. Its composition is shown in Fig. 1F, showing the phase is a combination of a linear phase and a cubic phase.

In this experiment, ST-Airy rings wavepacket is generated by a spatiotemporal wavepacket shaping and measuring system based on a Mach-Zehnder interferometer [30]. In such a configuration, the SLM can modulate the Fourier phase $\phi(k_x,\omega)$. Using this setup, we generate the ST Airy rings wavepacket and retrieve its three-dimensional intensity and phase information of the generated "object" light [30]. Figure 2A and B shows the measurement results for the generated ST-Airy rings wavepacket. The wavepacket is generated at the back focal plane of the Fourier lens and we mark this location as $z = 0$. The generated wavepacket has more than 8 ST-Airy rings. When such a wavepacket propagates dispersively, the joint effect of dispersion and diffraction leads to the spatiotemporal aufo-focusing of the wavepacket. In the experiment, we let the wavepacket propagate in a "virtual" dispersive medium with a dispersion coefficient $\beta_2$ of $9.5\times10^4$ fs$^2$/mm. The second and third column in Fig 2A and B show the abrupt auto-focusing effect of the wavepacket. After 275 mm, the ST-Airy rings wavepacket reaches the focus. At the focus, the central lobe of the wavepacket has a spatial width of 110 μm and a temporal width of 320 fs.

We also simulate the auto-focusing of the ST-Airy rings wavepacket with the same experimental parameters. The simulation results are shown in Fig. 2C. The simulation has a decent agreement with the experiment. We also calculate the contrast of central lobe intensity along the propagation distance. The results are shown in Fig. 2D. Triangle symbols in this figure comes from the experiments. The intensity contrast is also compared with with conventional Gaussian-Gaussian wavepacket. The ST-Airy rings wavepacket has a much more abrupt focus that can be useful for applications where a fast increasing of the light intensity and a shallow depth of focus is needed, for example, nonlinear microscopy and multi-photon 3D printing.

Beside aufo-focusing, Airy wave is also well known to have self-healing property. To demonstrate it, we put a 125-μm-wide fiber lead in the optical path slightly before the back focus of the Fourier lens. The ST-Airy rings wavepacket is then partially blocked. The left column in Fig 3A and B shows the measured ST-Airy rings after the spatial blockage. Upon dispersive propagation, the "injured" ST-Airy rings wavepacket can self-heal, and, in the same time, auto-focus spatiotemporally. The middle and right columns in Fig. 3A and B show the measurement results for this process. To compare, Figure 3C shows the simulation results under the same experimental condition. At about 285 mm after the fiber lead blockage, the wavepacket can self-heal and auto-focus. The resulting focal spot is about 200 μm in width and 270 fs in pulse duration. The results show that the wavepacket has a robust auto-focusing capability even during the presence of spatiotemporal perturbation.

Besides auto-focusing and self-healing, ST-Airy rings wavepacket can carry a transverse photonic orbital angular momentum (OAM) when an extra vortex phase is added to the Fourier phase applied on the SLM. Figure 4 shows the measurement of the ST-Airy vortex wavepacket and its spatiotemporal self-accelerating process. Due to the spatiotemporal vortex phase, the wavepacket cannot be focused. Consequently, this forms a tight spatiotemporal confinement of the transverse OAM (t-OAM) and it also has a large intensity gradiant. The centeral ring structure has a temporal spread of <400 fs and a spatial spread of <180 μm. Currently, this spread is limited by instrument resolution and other experimental conditions. Having a tight confinement of t-OAM may facilitate studying light and matter interaction that involves t-OAM.

**Discussion**

In conclusion, we present the first experimental demonstration of a spatiotemporal Airy rings wavepacket exhibiting an Airy function distribution in the radial direction in the spatiotemporal domain. This wavepacket exhibits self-accelerating dynamics governed by the interplay of dispersion and diffraction, leading to an abrupt autofocusing effect with a sharp intensity peak. Such a property holds promise for applications in nonlinear microscopy and multi-photon 3D printing. Notably, the autofocusing behavior remains robust against perturbations and obstructions, enhancing its applicability in practical scenarios. Additionally, we demonstrate that the wavepacket can carry transverse orbital angular momentum and confine it within a compact spatiotemporal volume. These unique features open new avenues for leveraging structured wavepackets in diverse optical applications, from advanced imaging to ultrafast laser processing.

**Materials and Methods**

### Phase diagram for generating ST Airy rings

The phase used for generating ST-Airy rings wavepacket can be mathematically expressed by a Fourier transformation from the spatial frequency and spectral domain to the real spatiotemporal domain [20],

$$G(\rho, \phi) = F.T.\{g_R(r)\exp(-il\theta)\} = 2\pi(-i)^l \exp(-il\phi)H_l\{g_R(r)\}, \quad (1)$$

where $F.T.$ stands for the Fourier transformation, $H_l$ stands for the $l$-th order Hankel transformation, $(\rho, \phi)$ is the polar coordinate in the real $(x, T)$ space, $(r, \theta)$ is the polar coordinate in the Fourier $(k_x, \omega)$ space. The spectral field in the Fourier space $g_R(r)$ is modulated by the spatial light modulator (SLM). The modulated Fourier field in the radial direction can be written as

$$g_R(r) = g_{R,0}(r) \cdot \exp(i\varphi_{\text{SLM}}(r)). \quad (2)$$

where $g_{R,0}(r)$ is the input field and $\varphi_{\text{SLM}}(r)$ stands for the applied SLM phase. At far field, under zero topological charge ($l = 0$), the transformation reduces to a zeroth-order Hankel transform,

$$H_l(g_R(r)) = \int_0^{+\infty} rg_R(r)J_l(2\pi\rho r)dr = \int_0^{+\infty} rg_{R,0}(r) \cdot J_0(2\pi\rho r) \cdot \exp(i\varphi_{\text{SLM}}(r))\,dr. \quad (3)$$

To generate ST-Airy rings wavepacket, the SLM phase $\varphi_{\text{SLM}}(r)$ is a superposition of a linear term and a cubic term (Fig. 1(f)), written as

$$\varphi_{\text{SLM}}(r) = a_{\text{linear}} \cdot r + a_{\text{third}} \cdot (r - r_0)^3. \quad (4)$$

Here, the linear phase term causes the radial shifting in real space and the cubic phase term generates the radial Airy wave. Equation (3) can be directly compared with the generation of Airy wave via Fourier transform. In Eqn. (3), $rg_{R,0}(r)$ is the spectral envelope term, $J_0(2\pi\rho r)$ approximates the sinusoid oscillation, and $\exp(i\varphi_{\text{SLM}}(r))$ has the third-order phase. Considering the bandwidth of the spectral field is finite so that Eqn. (4) can approximate the Fourier transform, we can use this pure radial phase modulation described by Eqn. (4) to generate the spatiotemporal (ST) Airy rings wavepacket using the spatiotemporal pulse shaper [29].

Experimental configuration

Figure 5 illustrates the experimental setup. A home-built fiber laser, as the master laser, is split into the "object" light and the "probe" light by a beamspliter (BS1 in the figure). In one arm, the "object" light passes a 4-f spatiotemporal pulse shaper constituting a diffraction grating, a cylindrical lens, and a liquid crystal spatial light modulator (SLM). After exiting the spatiotemporal pulse shaper, the light is focused by a Fourier lens with a focal length of 400 mm. In such a configuration, the SLM can modulate the Fourier phase $\phi(k_x,\omega)$. In the other arm, the "probe" light is compressed by a grating compressor and it is re-combined with the "object" light with a controllable delay at a CCD camera. By scanning the delay between the "object" and the "probe", we can record their interference fringes and then retrieve the three-dimensional intensity and phase information of the generated "object" light [30].

## Acknowledgements

**Funding:**

We acknowledge financial support from National Natural Science Foundation of China (NSFC) [Grant Nos. 12434012 (Q.Z.) and 12474336 (Q.C.)], the Shanghai Science and Technology Committee [Grant Nos. 24JD1402600 (Q.Z.) and 24QA2705800 (Q.C.)], National Research Foundation of Korea (NRF) funded by the Korea government (MSIT) [Grant No. 2022R1A2C1091890], and Global - Learning & Academic research institution for Master's·PhD students, and Postdocs (LAMP) Program of the National Research Foundation of Korea(NRF) grant funded by the Ministry of Education [No. RS-2023-00301938]. Q.Z. also acknowledges support by the Key Project of Westlake Institute for Optoelectronics [Grant No. 2023GD007].

**Author contributions:**

Conceptualization: AC, QC

Methodology: XLS, QC

Investigation: XLS

Visualization: XLS

Supervision: AC, QC, QWZ

Writing—original draft: XLS, QC

Writing—review & editing: AC, QC, QWZ

**Figures and Tables**

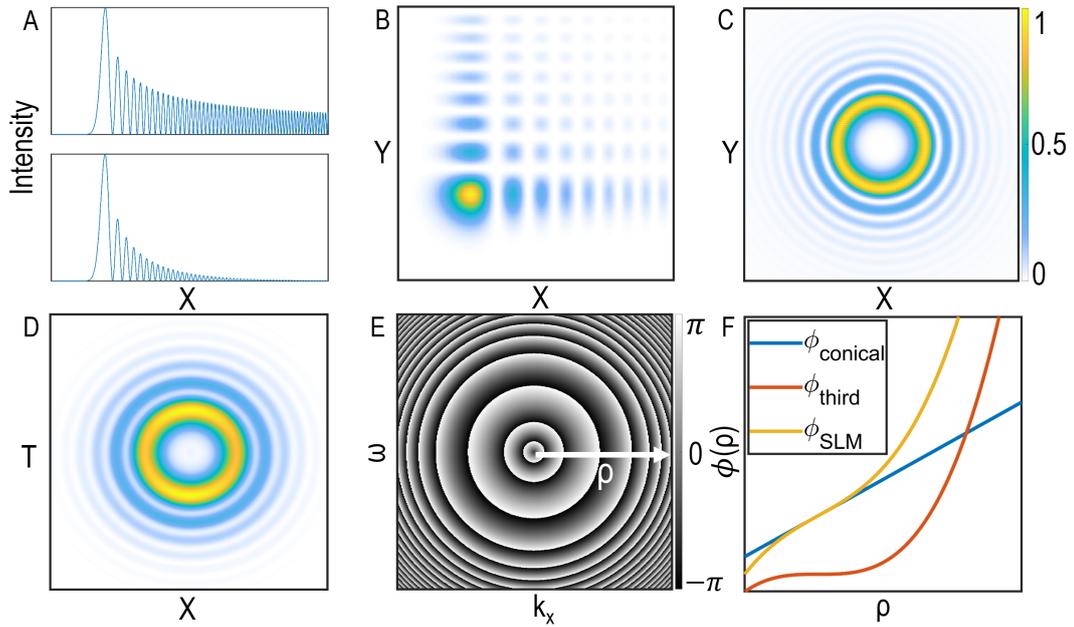

**Figure 1. Spaitiotemporal Airy rings wavepacket.** (A) Perfect 1D Airy wave (upper) and truncated 1D Airy wave with finite energy (lower). (B) 2D Airy-Airy beam. (C) Radially symmetric Airy rings beam. (D) Spatiotemporal (ST) Airy rings wavepacket. (E) Fourier phase for generating ST-Airy rings wavepacket $\phi(\omega, k_x)$. (F) Decomposition of the Fourier phase $\phi(\rho_{\omega-k_x})$ in the radial direction.

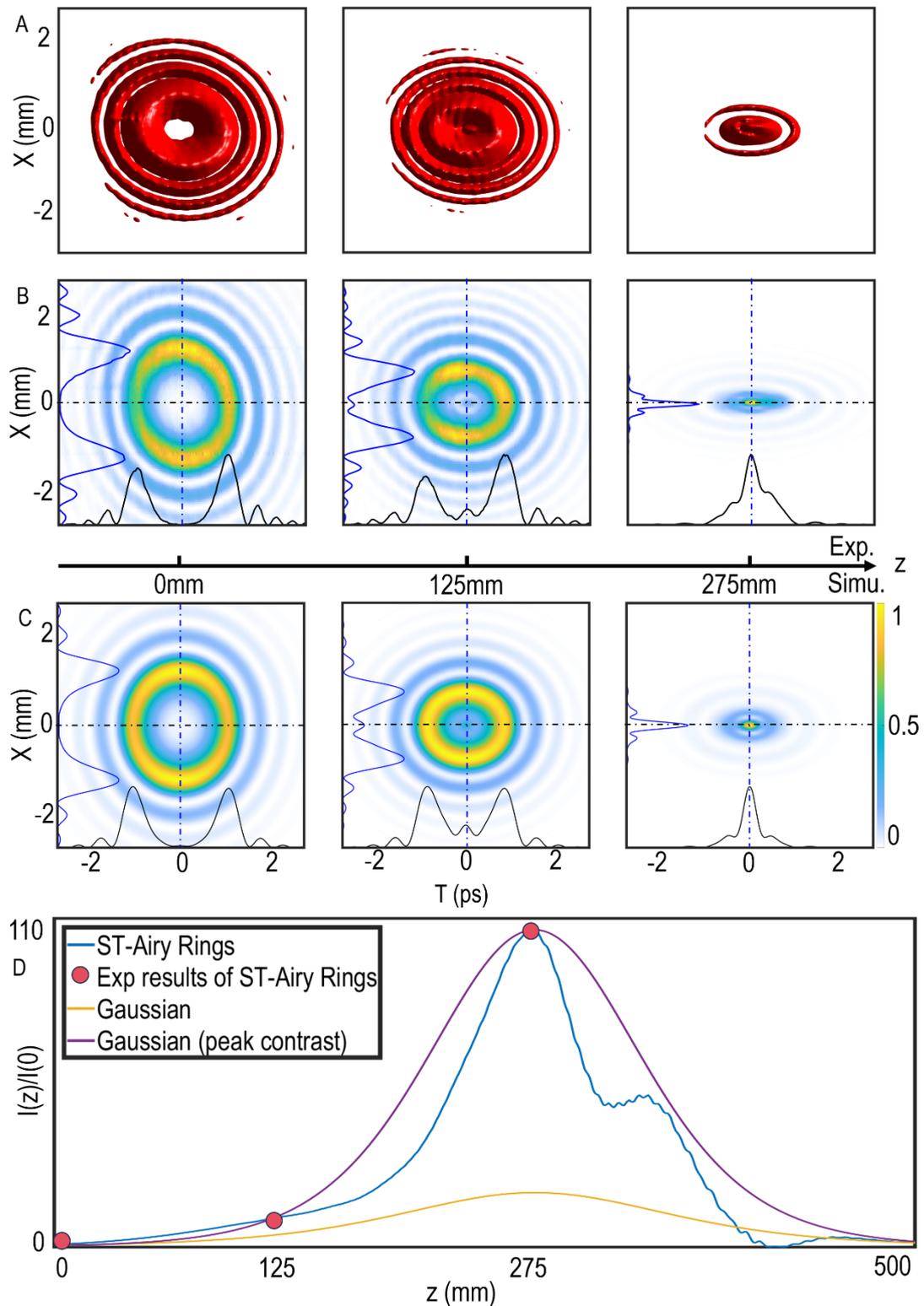

**Figure 2 Abrupt spatiotemporal auto-focusing of ST-Airy rings wavepacket.** (A) Three-dimensional intensity isosurface of ST-Airy rings wavepacket and its auto-focusing process. (B) Spatiotemporal intensity profile of the wavepacket during auto-focusing. (C) Simulation of ST-Airy rings wavepacket. (D) Comparison of the intensity contrast between ST-Airy rings wavepacket and conventional Gaussian-Gaussian wavepacket during focusing.

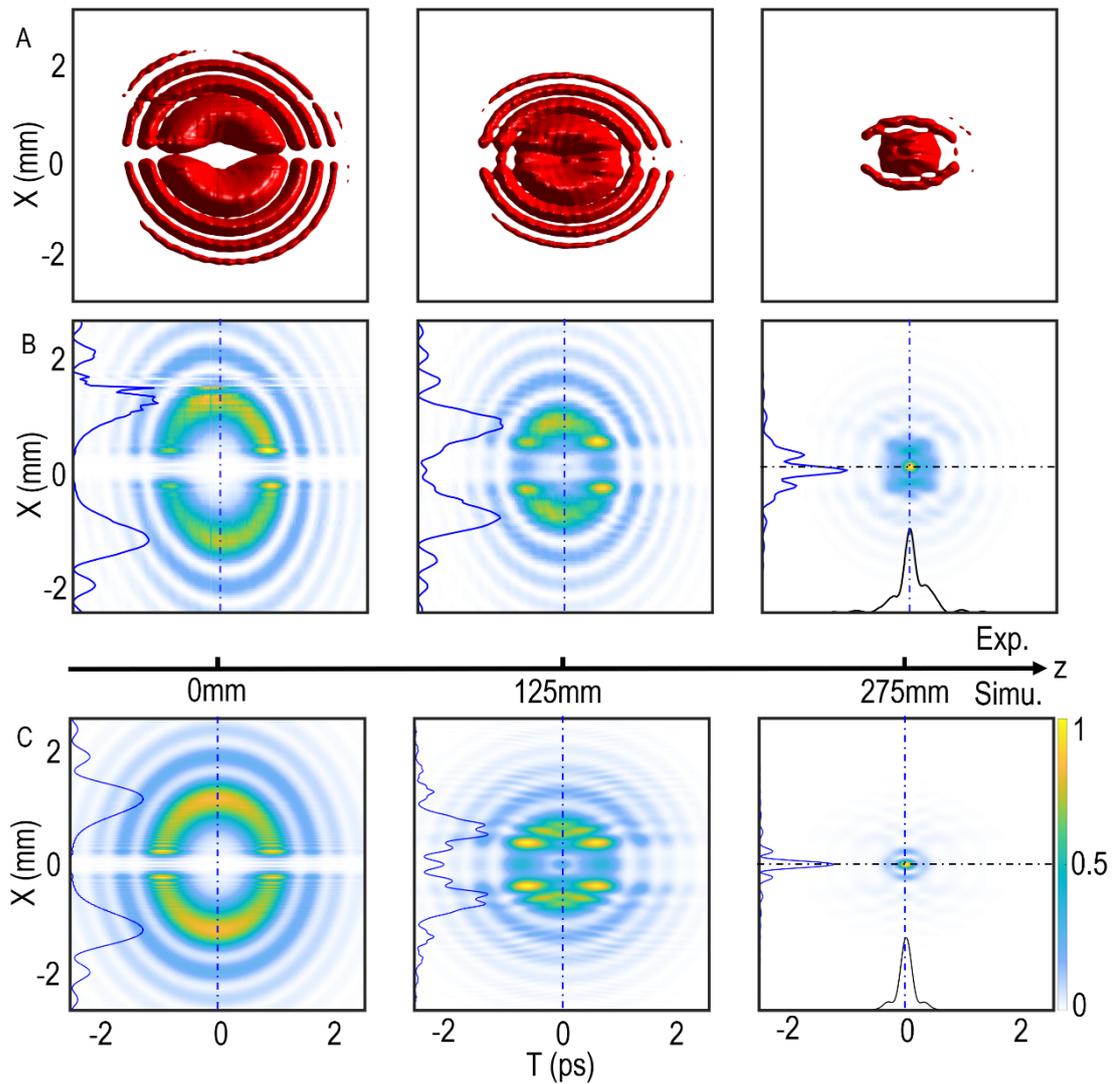

**Figure 3 Spatiotemporal auto-focusing of ST-Airy rings wavepacket after a spatial blockage.** The spatial blockage is a 125-μm-wide fiber lead. (A) Three-dimensional intensity isosurface of ST-Airy rings wavepacket after being partial blocked and then auto-focusing without being perturbed. (B) Spatiotemporal intensity profile. At focus, the central lobe of the wavepacket has a spatial width of 200 μm and a temporal width of 270 fs; (C) Simulation results for the ST-Airy rings wavepacket after the spatial blockage.

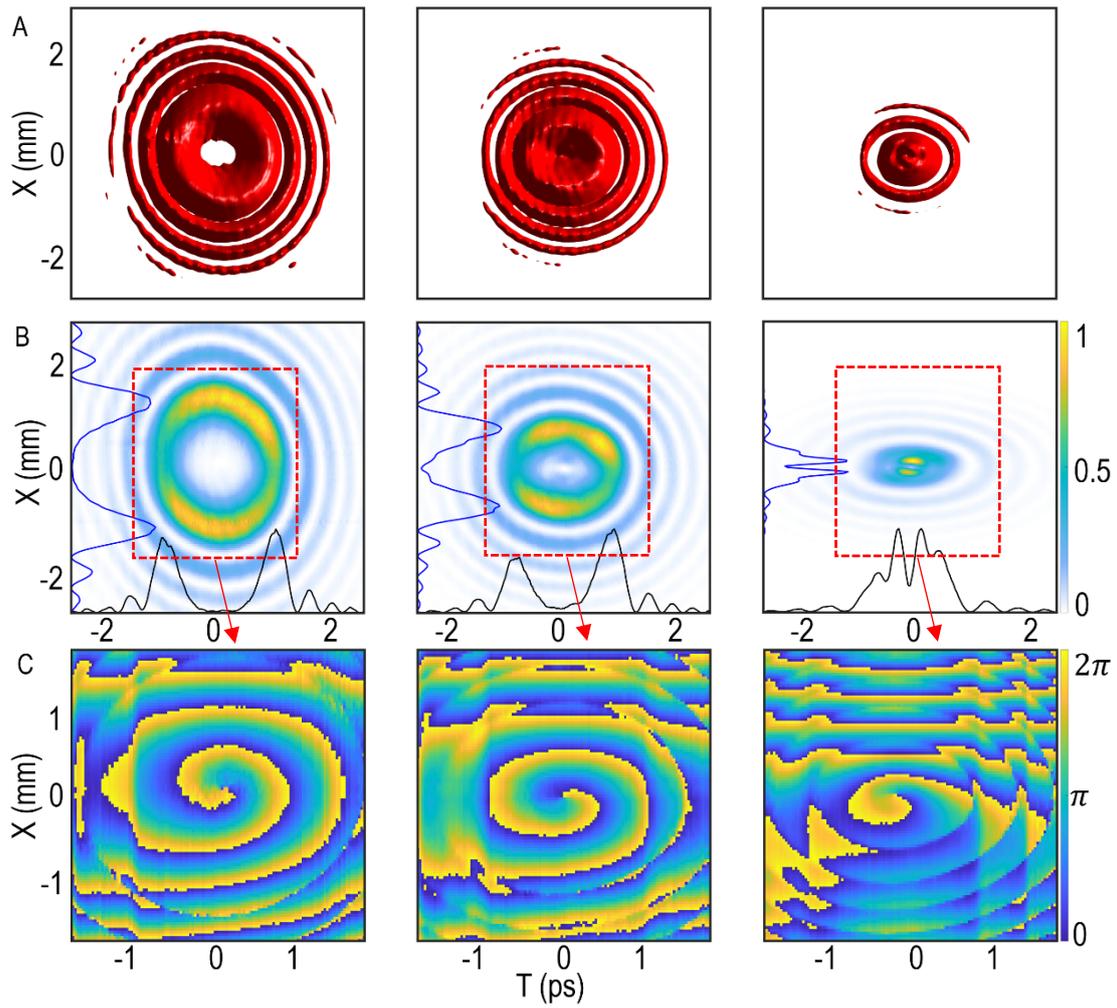

**Figure 4 Evolution of ST-Airy vortex rings wavepacket during dispersive propagation.**
(A) Three-dimensional intensity isosurface of ST-Airy vortex rings wavepacket carrying transverse OAM. (B) Spatiotemporal intensity profile. In the right-most figure, the transverse OAM is tightly confined within about 180 μm and 400 fs; (C) Spatiotemporal phase profile of the wavepacket.

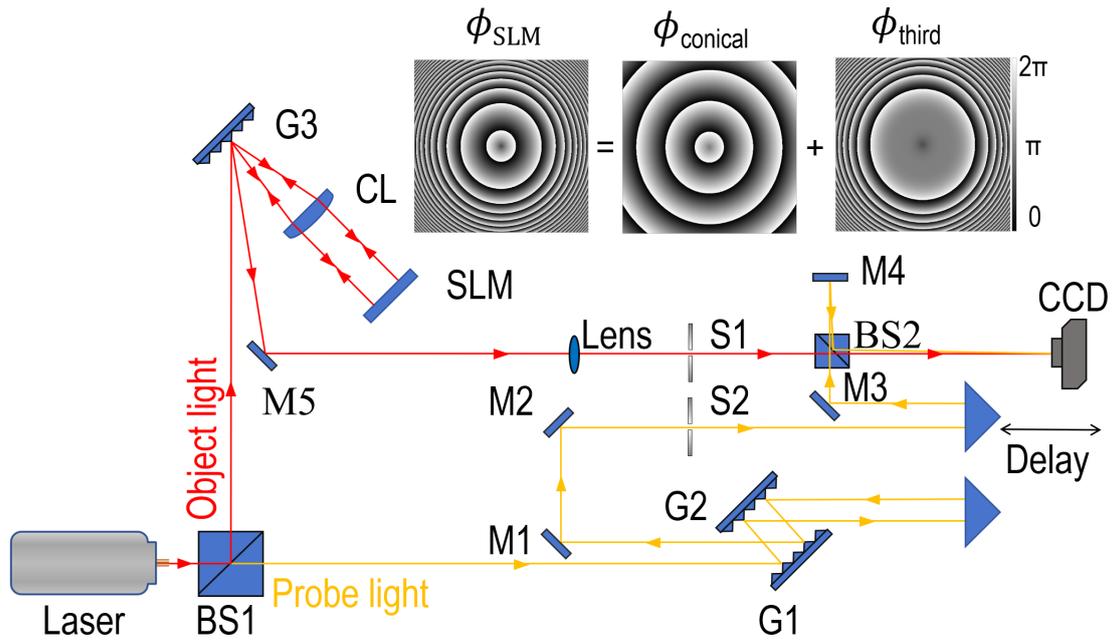

**Figure 5 Experimental setup for generating and measuring spatiotemporal Airy rings wavepacket.** The setup has a Mach-Zendher interferometric schematic that is similar with the one in [25]. M: mirror; BS: beamsplitter; G: grating; CL: cylindrical lens; SLM: spatial light modulator; S: shutter;